\documentclass[aps, prl, twocolumn, superscriptaddress, nofootinbib]{revtex4}
\usepackage{graphicx}
\usepackage{dcolumn}
\usepackage{amssymb,amsmath,amsthm,mathrsfs}
\usepackage{epstopdf}

\begin{document}

\newcommand{\Eq}[1]{Eq. \ref{eqn:#1}}
\newcommand{\Fig}[1]{Fig. \ref{fig:#1}}
\newcommand{\Sec}[1]{Sec. \ref{sec:#1}}

\newcommand{\PHI}{\phi}
\newcommand{\vect}[1]{\mathbf{#1}}
\newcommand{\Del}{\nabla}
\newcommand{\unit}[1]{\mathrm{#1}}
\newcommand{\x}{\vect{x}}
\newcommand{\ScS}{\scriptstyle}
\newcommand{\ScScS}{\scriptscriptstyle}
\newcommand{\xplus}[1]{\vect{x}\!\ScScS{+}\!\ScS\vect{#1}}
\newcommand{\xminus}[1]{\vect{x}\!\ScScS{-}\!\ScS\vect{#1}}
\newcommand{\diff}{\mathrm{d}}

\newcommand{\be}{\begin{equation}}
\newcommand{\ee}{\end{equation}}
\newcommand{\bea}{\begin{eqnarray}}
\newcommand{\eea}{\end{eqnarray}}
\newcommand{\vu}{{\mathbf u}}
\newcommand{\ve}{{\mathbf e}}
\newcommand{\vk}{{\mathbf k}}
\newcommand{\vx}{{\mathbf x}}
\newcommand{\vy}{{\mathbf y}}
\newcommand{\bx}{{\bf x}}
\newcommand{\bk}{{\bf k}}
\newcommand{\br}{{\bf r}}

\newcommand{\uden}{\underset{\widetilde{}}}
\newcommand{\den}{\overset{\widetilde{}}}
\newcommand{\denep}{\underset{\widetilde{}}{\epsilon}}

\newcommand{\nn}{\nonumber \\}
\newcommand{\dd}{\diff}
\newcommand{\fr}{\frac}
\newcommand{\del}{\partial}
\newcommand{\eps}{\epsilon}
\newcommand\CS{\mathcal{C}}

\def\be{\begin{equation}}
\def\ee{\end{equation}}
\def\ben{\begin{equation*}}
\def\een{\end{equation*}}
\def\bea{\begin{eqnarray}}
\def\eea{\end{eqnarray}}
\def\bal{\begin{align}}
\def\eal{\end{align}}

\def\TT{{\rm TT}}
\def\GW{{_{\rm GW}}}


\title{Anisotropies in the Gravitational Wave Background from Preheating}

\newcommand{\addressImperial}{Theoretical Physics, Blackett Laboratory, Imperial College, London, SW7 2AZ, United Kingdom}

\author{Laura Bethke}
\affiliation{\addressImperial}

\author {Daniel G. Figueroa}
\affiliation{D\'epartement de Physique Th\'eorique and Center for Astroparticle Physics, Universit\'e de Gen\`eve, 24 quai Ernest Ansermet, CH--1211 Gen\`eve 4, Switzerland}

\author{Arttu Rajantie}
\affiliation{\addressImperial}


\date{\today}

\begin{abstract}
We investigate the anisotropies in the gravitational wave (GW) background produced at preheating after inflation. Using lattice field theory simulations of a massless preheating model, we show that the GW amplitude depends sensitively on the value of the decay product field $\chi$ coupled to the inflaton $\phi$, with the only requisite that $\chi$ is light during inflation. We find a strong anisotropy in the amplitude of the GW background on large angular scales, the details of which strongly depend on the reheating dynamics. We expect similar conclusions for a wide class of inflationary models with light scalar fields. If future direct detection GW experiments are capable of detecting the GW produced by preheating, they should also be able to detect this effect. This could eventually provide a powerful way to distinguish between different inflationary and preheating scenarios. 
\end{abstract}

 \pacs{To be done}

\maketitle



The last two decades have seen enormous progress in cosmology, driven mainly by measurements of the temperature anisotropies of the cosmic microwave background (CMB) radiation, and culminating in the recent publication of the data from the Planck satellite~\cite{Planck}. In principle, gravitational waves (GW) have the potential to become an equally important source of information about the early universe. Their great advantage is that while the universe was opaque to electromagnetic radiation for the first $\sim 400.000$ years, GW were able to travel freely through space ever since their emission, and therefore they can carry information to us from the earliest moments of the universe~\cite{Maggiore}, close to the Big Bang. Possible sources of GW in the early universe include quantum fluctuations during inflation~\cite{StarobinskyGW}, non-equilibrium phenomena after inflation~\cite{GWpostInflation,preheatingGW}, and cosmic defects~\cite{GWdefects}.

One particularly interesting source of GW is reheating, the transition from inflation to a radiation phase. During this process, the inflaton field, which was driving the inflation, decays into other fields. In many models, reheating begins with an initial non-equilibrium stage known as preheating~\cite{preheating}, which is much more sensitive to the details of the model than the period of inflation itself. In particular, preheating generates a sub-horizon background of GW with a significant amplitude~\cite{preheatingGW}, whose frequency depends on the energy scale of inflation, being typically in the MHz-GHz range. This is too high to be observed with existing or planned large GW detectors such as Advanced LIGO, but prototype detectors for 100 MHz GW have been built~\cite{MHz}.

In the existing literature, the focus has been on determining the spectrum of the GW, but the purpose of this letter is to show that under certain relatively general conditions, the GW amplitude can depend on position on cosmological scales. This means that the observed GW background will be anisotropic on large angular scales. Such anisotropy should be potentially observable with any GW experiment that is capable of detecting the GW background, and might provide a powerful new way to constrain models of inflation and preheating.

The anisotropy arises in models with a second light scalar field $\chi$, in addition to the inflaton field. By light we mean that its mass is less than the Hubble rate $H_*$ during inflation, thanks to which it should develop an almost Gaussian scale-invariant spectrum of fluctuations during inflation (similarly to the inflaton), with power spectrum ${\cal P}_\chi 
\approx {H_*^2/4\pi^2}$ \cite{LiddleLyth}. When inflation ends, the details of preheating in each Hubble volume depend on the local value of the $\chi$ field, which we denote by $\chi_{\rm i}$. In particular, this will also be true for the energy density $\rho_\GW$ of the GW produced within each Hubble volume, which therefore becomes a local function of $\chi_{\rm i}$. That is, normalizing the GW energy density to the critical energy density $\rho_c$ today, $\Omega_\GW \equiv \rho_\GW/\rho_c$, we expect $\Omega_{\GW}=\Omega_{\GW}(\chi_{\rm i})$. 

The GW background from preheating originated from a comoving spherical shell of radius $R\sim 1/H_0$, with $H_0$ the Hubble rate today. This surface encompasses a very large number of preheating Hubble patches. The GW background we could detect in a given direction ${\hat n}$, was then produced within one of these patches centered at ${\bf r}=R{\hat n}$. We expect therefore that its amplitude should depend on the value of $\chi_i$ within the preheating Hubble volume at ${\br}=R{\hat n}$. Because $\chi_i$ has fluctuations on cosmological scales, so does $\Omega_{\GW}$, and therefore the observed GW amplitude will depend on the direction ${\hat n}$, i.e.~$\Omega_{\GW}({\hat n})= \Omega_{\GW}(\chi_{\rm i}(R{\hat n}))$.  

The details of the anisotropy are determined by how the GW energy density depends on $\chi_{\rm i}$, i.e.~by the function $\Omega_{\GW}(\chi_{\rm i})$. We will present a lattice calculation of $\Omega_\GW(\chi_{\rm i})$ for a massless preheating model shortly, but to illustrate the effect let us first assume a linear dependence as
\be
\label{equ:lindep}
\Omega_\GW(\chi_{\rm i})=c_0+c_1(\delta\chi_{\rm i}/H_*),
\ee
where $c_0$ and $c_1$ are dimensionless constants, and $\delta\chi_{\rm i}=\chi_{\rm i}-\overline{\chi}_{\rm i}$, with $\overline{\chi}_{\rm i}$ the average value over the currently observable universe. In this case the relative fluctuations of the GW energy density are $\delta\Omega_\GW \equiv (\Omega_\GW/\overline{\Omega}_\GW-1)=(c_1/c_0)(\delta\chi_{\rm i}/H_*)$. Like $\delta\chi_i$, these fluctuations are nearly Gaussian and scale-invariant, and their power spectrum is
${\cal P}_{\GW}=(c_1/c_0)^2({\cal P}_{\chi}/H_*^2)= (c_1/2\pi c_0)^2.$ The angular power spectrum of the relative fluctuations on large scales (i.e.~small multipole $l$), is given by
\be
\label{equ:Cl}
l(l+1)C_l=\frac{\pi}{2}{\cal P}_{\GW}=\frac{1}{8\pi}{c_1^2\over c_0^2},
\ee
in analogy with the Sachs-Wolfe plateau~\cite{LiddleLyth}.

In practice, $\Omega_\GW(\chi_{\rm i})$ will not be linear and therefore the calculation of the effect is more complicated~\cite{Suyama:2013dqa}. 
In order to quantify the anisotropies on large angular scales, we compute the two-point correlation function of the GW energy density originated at two points $\bx$ and ${\bf y}$. 
The joint probability distribution for the field values $\chi_{\bf x}=\chi_{\rm i}({\bf x})$ and $\chi_{\bf y}=\chi_{\rm i}({\bf y})$ at these points is
\be
P(\chi_{\bf x},\chi_{\bf y})\propto 
\exp\left[-\frac{\sigma_\chi^2(\delta\chi_{\bf x}^2+\delta\chi_{\bf y}^2)-2G_{{\bf x},{\bf y}}\delta\chi_{\bf x}\delta\chi_{\bf y}}{2(\sigma_\chi^4-G_{{\bf x},{\bf y}}^2)}
\right],
\ee
where $\delta\chi=\chi_{\rm i}-\overline{\chi}_{\rm i}$,  $G_{{\bf x},{\bf y}}\equiv\langle\delta\chi_{\rm i}({\bf x})\delta\chi_{\rm i}({\bf y})\rangle$ is the field correlator,
and $\sigma_\chi^2=G_{{\bf x},{\bf x}}=\langle\delta\chi^2\rangle$ is the field variance.
The GW energy density correlator is then given by
\bea
&&\left\langle \Omega_{\GW}(\bx)\Omega_{\GW}({\bf y}) \right\rangle \equiv \nonumber\\ 
&&~~~~~\int d\chi_{\bf x}d\chi_{\bf y}P(\chi_{\bf x},\chi_{\bf y}) \Omega_{\GW}(\chi_{\bf x})\Omega_{\GW}(\chi_{\bf y}).\label{eq:GWcorr}
\eea
The analysis simplifies if the correlator is well approximated by its linear Taylor expansion in powers of the field correlator $G_{{\bf x},{\bf y}}$, 
\be
\label{equ:linexp}
\left\langle \Omega_{\GW}({\bf x})\Omega_{\GW}({\bf y}) \right\rangle \simeq 
\langle\Omega_\GW\rangle^2+\frac{\left\langle
\delta\chi_{\rm i}\Omega_\GW(\chi_{\rm i})
\right\rangle^2
}{\sigma_\chi^4}
G_{{\bf x},{\bf y}},
\ee
where the expectation values on the right hand side are computed with the single-point probability distribution
\be
\label{equ:singleP}
P(\chi_{\rm i})\propto\exp\lbrace-(\chi_{\rm i}-\overline{\chi}_{\rm i})^2/2\sigma_\chi^2\rbrace. 
\ee
Then, the GW energy density correlator coincides with the linear ansatz (\ref{equ:lindep})
with coefficients $c_0=\langle\Omega_\GW\rangle$ and 
$
c_1=H_*\left\langle
\delta\chi_{\rm i}\Omega_\GW(\chi_{\rm i})
\right\rangle/\sigma_\chi^2.
$
The angular power spectrum is therefore given by Eq.~(\ref{equ:Cl}),
\be
\label{equ:Cl2}
l(l+1)C_l=\frac{H_*^2}{8\pi}\frac{\langle\delta\chi_{\rm i}\Omega_\GW(\chi_{\rm i})\rangle^2}{\sigma_\chi^4\langle\Omega_\GW\rangle^2}.
\ee

We have calculated this explicitly in the  massless preheating model. It is one of the easiest models to analyze numerically, and it has been extensively studied analytically and numerically \cite{Kofman2, prokopec}. It describes the inflaton $\phi$, with a chaotic quartic potential during inflation, coupled to a scalar $\chi$ as 
\be \label{masslesspre} V(\phi,\chi) = \frac{\lambda}{4} \phi^4 + \frac{1}{2} g^2 \phi^2 \chi^2, \ee
with $\lambda=9\times 10^{-14}$ as fixed by the amplitude of the CMB anisotropies. This model is conformally invariant, which makes it very convenient for lattice simulations, since comoving physical scales will remain within the lattice volume throughout the simulation. Although massless preheating has been nearly ruled out by observations \cite{komatsu}, we use it for its numerical convenience. We expect our conclusions to be valid for more realistic models as well.

During inflation $\phi$ is large, $\phi\gtrsim M_{\rm Pl}$, and $\chi$ is assumed to be very small.
When inflation ends,  $\phi$ starts oscillating around its minimum at $\phi = 0$. These oscillations induce an instability in the fluctuations of the field $\chi$. The amplitude of the $\chi$ modes initially grows exponentially fast for certain comoving momentum bands as $\chi_k(t) \propto e^{\mu_k\eta(t)}$, where $d\eta = dt/a(t)$ is the conformal time. This effect is known as parametric resonance~\cite{preheating} and has been studied analytically in great detail for this model in~\cite{Kofman2}. The unstable solutions where $\chi_k$ grows exponentially are characterized by an exponent $\mu_k$, which depends on the wavenumber $k$ and the ratio of coupling  constants $g^2/\lambda$. The band structure of this model is summarized in Fig.~4 of \cite{Kofman2}. We focused on the parameter choice $g^2/\lambda=2$, for which $\chi$ is light during inflation and parametric resonance is strongest for the longest wavelengths. Eventually the exponential growth ends when the linear approximation breaks down. 

The GW are primarily produced during the subsequent stage of non-linear non-equilibrium dynamics, which we study by using lattice field theory simulations. 
These consist of solving a discretized version of the scalar field equations of motion of the matter fields in the model,
\bea\label{eq:eomPhi}
\ddot \phi + 3H\dot \phi - \frac{1}{a^2}\nabla^2\phi + (\lambda\phi^2 + g^2\chi^2)\phi &=& 0,\\
\label{eq:eomChi}
\ddot \chi + 3H\dot \chi - \frac{1}{a^2}\nabla^2\chi + g^2\phi^2\chi &=& 0,
\eea
coupled to the Friedmann equation $3\ddot a/a = -4\pi(\rho + 3p)/3M_{\rm Pl}^2$ for the scale factor $a(t)$, sourced by the volume averaged pressure $p$ and energy density $\rho$ contributed by $\phi$ and $\chi$. The Planck mass is defined in terms of Newton's constant $G$ as $M_{\rm Pl}=G^{-1/2}$.
Our code was based on the publicly available MPI C/C++ ClusterEasy package~\cite{CLUSTEREASY}, which
solves the field equations with a second-order leap-frog integrator using periodic boundary conditions. 

To compute the GW spectrum  we followed the algorithm proposed in~\cite{Figueroa1}.
During preheating the system develops an anisotropic stress tensor $\Pi_{ij} = T_{ij} - pg_{ij}$, where $T_{ij}$ is the energy-momentum tensor. The transverse-traceless part of $\Pi_{ij}$ acts as a source for the tensor perturbation $h_{ij}$ representing GW. In practice, the GW source is given by $\Pi_{ij}^\TT = \lbrace\partial_i\chi\partial_j\chi + \partial_i\phi\partial_j\phi\rbrace^\TT$, where $\lbrace \cdots \rbrace^\TT$ stands for a projection of the transverse-traceless (TT) degrees of freedom, guaranteeing $\partial_i\Pi_{ij}^\TT = \Pi_{ii}^\TT = 0$. 
Therefore we solve a discretized version of the linearized Einstein equations for the tensor perturbations $h_{ij}$ in the presence of the non-linear field evolution determined by Eqs.~(\ref{eq:eomPhi}) and (\ref{eq:eomChi}), 
\bea
\label{eq:eomGW}
\ddot h_{ij} + 3H\dot h_{ij} - \frac{1}{a^2}\nabla^2h_{ij} = \frac{16\pi}{M_{\rm Pl}^2a^2} \Pi_{ij}^\TT(\phi,\chi).\eea
We verify the conditions $\partial_ih_{ij} = h_{
ii} = 0$ by choosing a lattice-based TT-projector associated to neutral lattice derivatives~\cite{Figueroa2}, but checked that the choice of projector did not influence our results.
Finally, we obtained the total energy density of GW within a volume $V$ and normalized to the critical energy density, as
\bea \Omega_\GW(t) = \frac{1}{\rho_c} \int \left(\frac{d\rho_\GW}{d\log k}\right) d\log k ~~~~~~~~\\
\frac{d\rho_\GW}{d\log k} \equiv \frac{k^3M_{\rm Pl}^2}{(4\pi)^3 V} \int {d\Omega_k\over 4\pi} \dot{h}_{ij}(t,k,\hat\vk)\dot{h}^{\ast}_{ij}(t,k,\hat\vk)  \label{rhoGW} 
\eea
For the discretized version of Eq.~(\ref{rhoGW}), see Eq.~(4.5) in~\cite{Figueroa2}. The GW amplitude grows with time, and eventually saturates, reaching a final amplitude at some time $t_f$, when the fields enter a turbulent regime. 
 
\begin{figure}[t]
\begin{center}
\includegraphics[width=8cm]{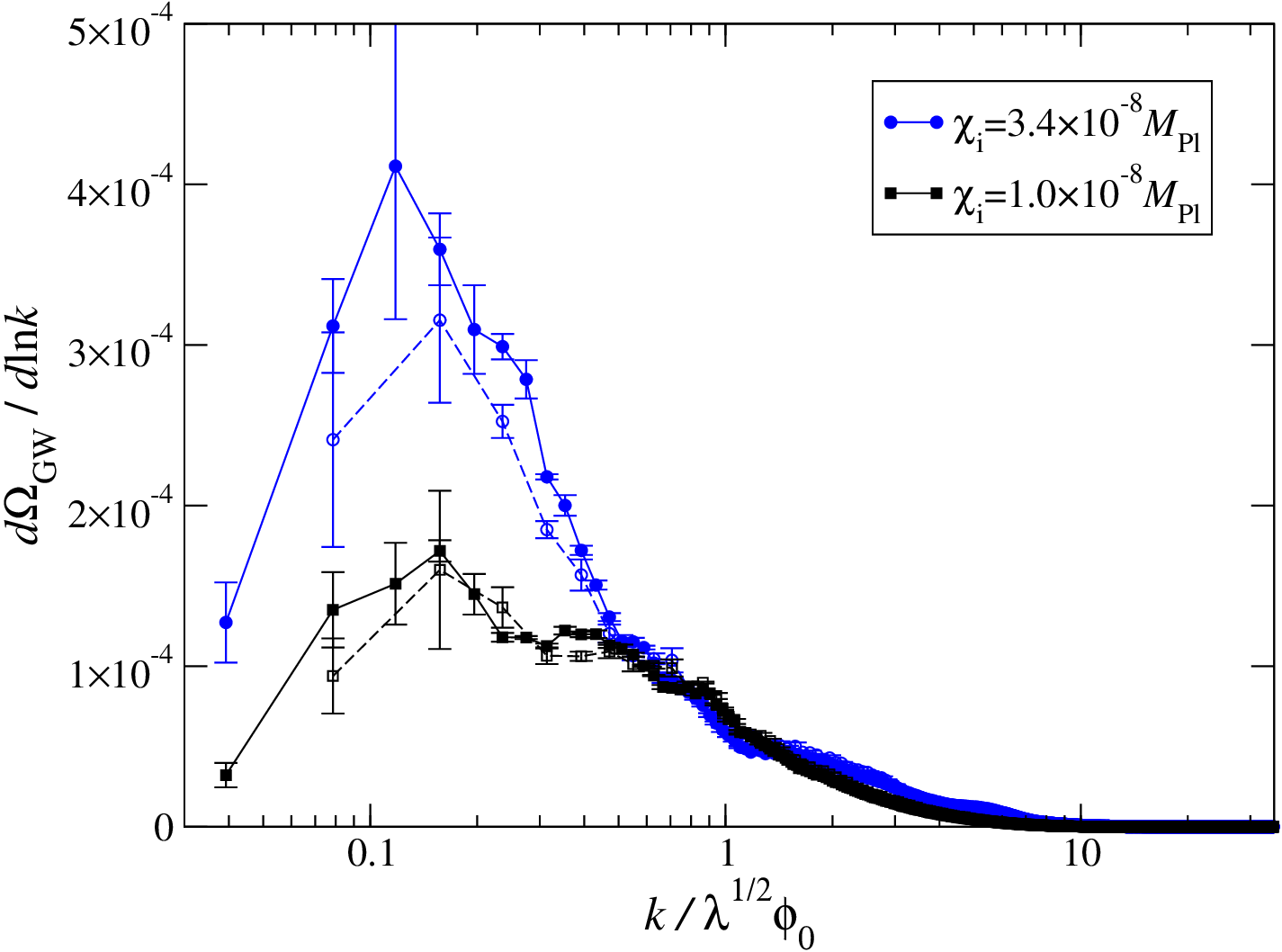}
\end{center}
\caption{Final spectrum of GW for 
$\chi_{\rm i} = 3.4\times 10^{-8}M_{\rm Pl}$ 
(upper, blue curve)
and 
$\chi_{\rm i}=1.0\times 10^{-8}M_{\rm Pl}$ 
(lower, black curves), 
averaged over five random realizations of inhomogeneous fluctuations. The solid curves are for $\tilde{L}=160$, $N=1024$, and the dashed curves for $\tilde{L}=80$, $N=512$. The area underneath corresponds to the total fractional GW energy density within a preheating Hubble domain.}
\label{fig:GWspectrum}
\end{figure}

As explained earlier, we want to determine how the final fractional GW energy density $\Omega_\GW(t_f)$ generated within a single Hubble region of volume $V = H_*^{-3}$ depends on the initial value $\chi_{\rm i}$ within same the region. Therefore, we repeated this calculation for a range of different $\chi_{\rm i}$. This is in contrast with most of the literature on massless preheating to date, in which $\chi_{\rm i}$ has been assumed to be zero. The main exception is the calculation curvature perturbations produced by preheating, which depends on the impact of $\chi_{\rm i}$ on the expansion rate \cite{rajantie}. These studies have demonstrated a chaotic behaviour which leads to non-Gaussian spikes in the curvature perturbation $\zeta$~\cite{Bond}.

The initial conditions for the fields $\phi$ and $\chi$ consisted of inhomogeneous Gaussian fluctuations mimicking vacuum quantum fluctuations \cite{khlebnikovtkachev}, superimposed on homogeneous background values $\phi_0=0.342M_{\rm Pl}$, $\chi_0=\chi_{\rm i}$.

Our simulations show the expected behavior of parametric resonance: the inflaton oscillates initially, transferring energy to the fluctuations in ${\chi}$ which grow until the system becomes non-linear. As the fluctuations grow, so does the gradient energy of the fields which leads to GW production. The field gradients eventually become an adiabatic slowly evolving function, which sets the end of production of GW. In practice, the GW spectrum continues to oscillate, and we therefore took a time average over several oscillations to obtain the final GW amplitude. Fig.~\ref{fig:GWspectrum} shows the shape of the GW spectrum for two initial field values $\chi_{\rm i}$. The dashed lines corresponds to our fiducial choice of lattice size $\widetilde L \equiv \sqrt{\lambda}\phi_0L = 80$ and lattice points per dimension $N = 512$, whereas the solid lines correspond to $\widetilde L=160$ and $N=1024$ [ensuring the same ultraviolet (UV) coverage]. For $\widetilde L=160$, one can clearly see a drop in the infrared (IR), which  shows that very long wavelength modes are not excited. In practice, the total integrated GW amplitude for the two volumes agree to better than $\sim 1\%$. Because $\widetilde L=160$ was computationally too expensive and it was not possible to capture both IR and UV behaviors sufficiently well with smaller lattices, we chose $\widetilde L=80, N = 512$ for our simulations.

\begin{figure}[t]
\begin{center}
\includegraphics[width=8cm]{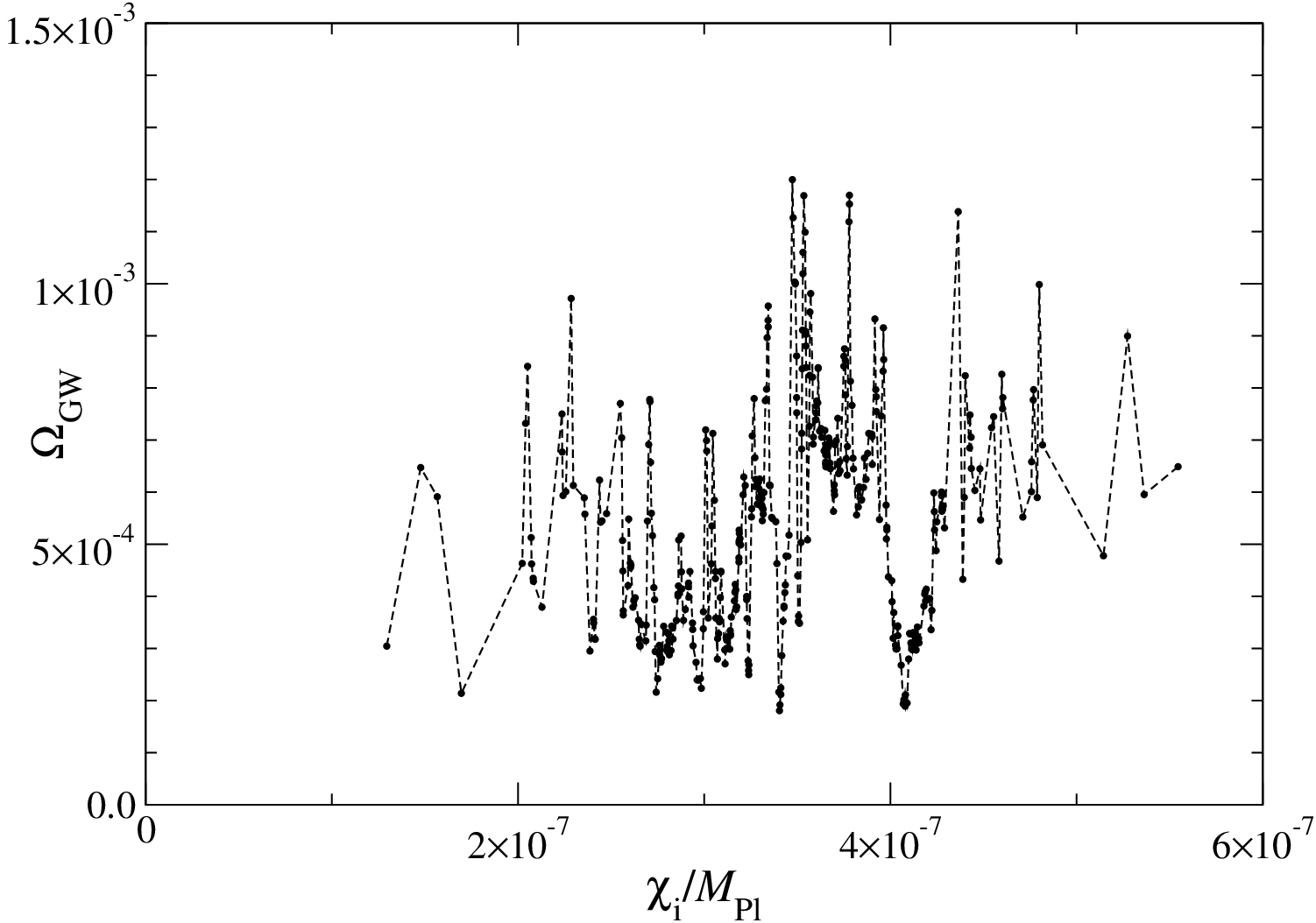}
\end{center}
\caption{$\Omega_\GW$ for our sample of initial field values $\chi_{\rm i}$.}
\label{fig:GWvsChi}
\end{figure}

To determine the anisotropy of the GW background, we need to compute the expectation values in Eq.~(\ref{equ:Cl2}) with  the Gaussian distribution (\ref{equ:singleP}), with variance
\be
\sigma_\chi^2=
\int_{a_0H_0}^{H_*} \frac{dk}{k}{\cal P}_\chi=\frac{H_*^2}{4\pi^2}N_{\rm CMB}
\approx
3.3\times 10^{-15}M_{\rm Pl}^2,
\ee
where we have used 
the 
Hubble rate at the end of inflation $H_{\ast}^2\approx 8\pi\lambda\phi_0^4/12 M_{\rm Pl}^2\approx 2.6\times 10^{-15}M_{\rm Pl}^2
$ and $N_{\rm CMB}\sim 50$ as the number of e-folds after the largest observable scales left the horizon. Because it is likely that inflation lasted more than 50 e-foldings, the scale invariant fluctuations of the $\chi$ field continue outside our current horizon, and therefore we have to allow for a non-zero average value $\overline{\chi}_{\rm i}$ over the currently observable universe, with variance $\langle{\overline\chi}_{\rm i}^2\rangle=(H_*^2/4\pi^2)(N_{\rm tot}-N_{\rm CMB})$, where $N_{\rm tot}$ is the total number of e-foldings. We treat $\overline{\chi}_{\rm i}$ as a free parameter with approximate magnitude $\overline{\chi}_{\rm i}\sim 10^{-7}M_{\rm Pl}$. 

We calculated the expectation values in Eq.~(\ref{equ:Cl2}) using the Monte Carlo method, generating ${\cal N}= 400$ random values $\chi_{\rm i}^j$, $j\in\{1,\ldots,{\cal N}\}$ 
from the Gaussian 
distribution (\ref{equ:singleP}), choosing the mean value $\overline\chi_{\rm i}=3.42\times 10^{-7}M_{\rm Pl}$. 
For each $\chi_{\rm i}^j$, we did one simulation run, measuring the GW energy density $\Omega_\GW(\chi_{\rm i}^j)$, see Fig.~\ref{fig:GWvsChi}.
As the plot shows, $\Omega_\GW$
is highly dependent on $\chi_{\rm i}$, varying by as much as a factor of five between nearby values, although there are some ranges of $\chi_{\rm i}$ where the dependence is much smoother. This irregular behavior is in line with the chaotic dynamics observed earlier~\cite{Bond}, but its amplitude is unexpectedly high.

\begin{figure}[t]
\begin{center}
\includegraphics[width=8cm]{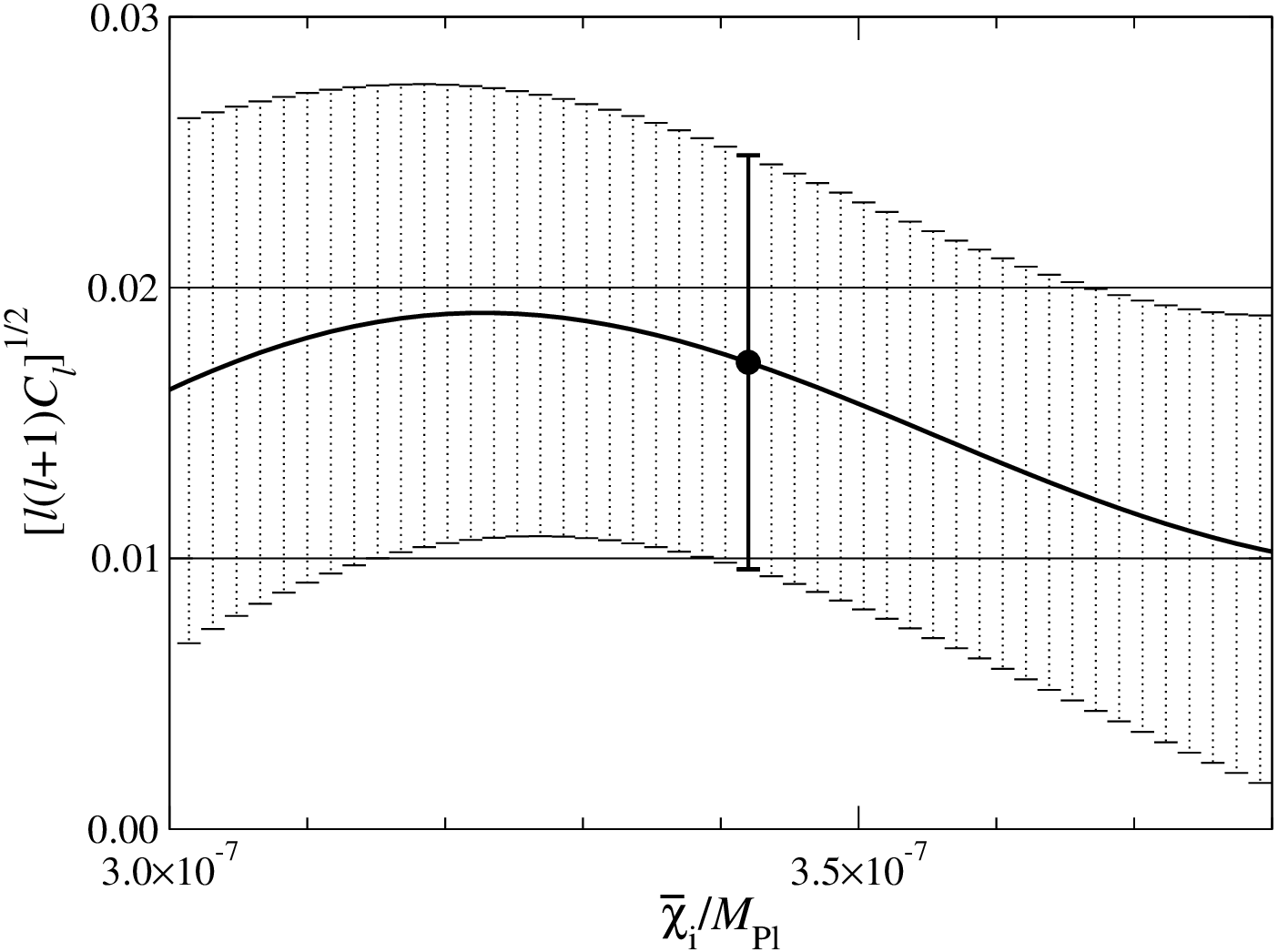}
\end{center}
\caption{
\label{fig:Clvschi}
The relative amplitude of the multipoles of the GW background as a function of the average field value $\overline{\chi}_{\rm i}$, calculated from Eq.~(\ref{equ:Cl2}). The black dot shows the amplitude for original mean value $\overline{\chi}_{\rm i}=3.42\times 10^{-7}M_{\rm Pl}$, and the curve shows values obtained by reweighting the same data.
}
\end{figure}

We computed the full correlator~(\ref{eq:GWcorr}) using the Monte Carlo data to confirm the validity of the linear expansion (\ref{equ:linexp}) and Eq.~(\ref{equ:Cl2}).
The expectation values in Eq.~(\ref{equ:Cl2}) are approximated by averages within our sample, 
\bea
\label{equ:averages}
\langle\Omega_\GW\rangle&\approx& \frac{1}{\cal N}\sum_j\Omega_\GW(\chi_{\rm i}^j),
\nonumber\\
\langle\delta\chi\Omega_\GW\rangle&\approx& \frac{1}{\cal N}\sum_j(\chi_{\rm i}^j-\overline{\chi}_{\rm i})\Omega_\GW(\chi_{\rm i}^j).
\eea
For $\overline\chi_{\rm i}=3.42\times 10^{-7}M_{\rm Pl}$, we obtained
$\langle\Omega_\GW\rangle= (5.45\pm 0.13)\times 10^{-4}$ and $\langle\delta\chi\Omega_\GW\rangle= (3.0\pm1.2)\times 10^{-12}M_{\rm Pl}$. Substituting these into Eq.~(\ref{equ:Cl2}) gives the amplitude of the relative fluctuations $\delta\Omega_\GW=(\Omega_{\GW}/\overline{\Omega}_\GW-1)$ as 
\be
\sqrt{l(l+1)C_l}=0.017\pm0.008,
\ee
where the errors are estimated by the bootstrap method.
We extended our results to nearby mean values $\overline{\chi}_{\rm i}'$
by reweighting our data, giving to each $\chi_{\rm i}^j$ the weight 
\be
r_j=\exp\left[-\frac{(\chi_{\rm i}^j-\overline\chi_{\rm i}')^2}{2\sigma_\chi^2}
+\frac{(\chi_{\rm i}^j-\overline\chi)^2}{2\sigma^2}\right]
\ee 
in the averages~(\ref{equ:averages}),
with $\overline\chi$ and $\sigma^2$ the actual numerical average and variance of our sample. The resulting amplitudes are shown in Fig.~\ref{fig:Clvschi}.
From these data we can conclude with some confidence that the relative amplitude of the anisotropies is above one per cent level in this case, much higher than the CMB anisotropies.

In summary, 
we have shown that the GW background from preheating is generally anisotropic on large angular scales if light scalar fields are present during inflation. We have used numerical lattice field theory simulations to demonstrate that in the massless preheating model the amplitude of these anisotropies is significant, at the $\sim 1 \%$ level. Obviously, measuring these anisotropies will be a great challenge even if the GW background itself is detected, but if achieved, it might provide detailed information about the microscopic physics of inflation.

The massless preheating model is exceptional because of its conformal invariance, and it is likely that in more realistic theories the GW energy density $\Omega_\GW$ is less sensitively dependent on the field value $\chi_{\rm i}$, like the curvature perturbation~\cite{Chambers:2009ki}. This does not necessarily mean that the amplitude of the anisotropies will be lower, because in the current case the highly irregular variation in Fig.~\ref{fig:GWvsChi} becomes largely averaged out. 
Our general conclusions are also not necessarily restricted to models with parametric resonance, and we would generally expect the GW background to be anisotropic on large angular scales in any model in which GW emission is affected by a light scalar field. 

In light of the recent discovery of a Higgs-like particle at the LHC~\cite{Aad:2012tfa}, it would be interesting to investigate this phenomenon in a realistic particle physics scenario, with the Standard Model Higgs field playing the role of the light scalar field.
This will be computationally more demanding and will require the inclusion of gauge fields~\cite{Figueroa3} and fermions~\cite{Figueroa4}.

The research was funded by STFC grants ST/J000353/1 and ST/F007027/1, and the Royal Society International Joint Project JP100273. DGF is supported by the Swiss National Science Foundation. The simulations were carried out using the COSMOS@DiRAC facility which is supported by STFC/DBIS UK.



\begin{thebibliography}{99}
\bibitem{Planck} 
  P.~A.~R.~Ade {\it et al.}  [ Planck Collaboration],
  arXiv:1303.5062 [astro-ph.CO].
\bibitem{Maggiore}
  M.~Maggiore,
  Phys.\ Rept.\  {\bf 331}, 283 (2000).
\bibitem{StarobinskyGW} 
  A.~A.~Starobinsky,
  JETP Lett.\  {\bf 30}, 682 (1979)
  [Pisma Zh.\ Eksp.\ Teor.\ Fiz.\  {\bf 30}, 719 (1979)].
\bibitem{GWpostInflation}
M.~Kamionkowski, A.~Kosowsky and M.~S.~Turner,
  Phys.\ Rev.\ D {\bf 49}, 2837 (1994)
  [astro-ph/9310044].
C.~Caprini, R.~Durrer and G.~Servant,
  JCAP {\bf 0912}, 024 (2009)
  [arXiv:0909.0622 [astro-ph.CO]]; 
 C.~Caprini, R.~Durrer, T.~Konstandin and G.~Servant,
  Phys.\ Rev.\ D {\bf 79}, 083519 (2009)
  [arXiv:0901.1661 [astro-ph.CO]].
 \bibitem{preheatingGW}
S.~Y.~Khlebnikov and I.~I.~Tkachev,
  Phys.\ Rev.\ D {\bf 56}, 653 (1997);
R.~Easther and E.~A.~Lim,
  JCAP {\bf 0604}, 010 (2006);
J.~Garc\'ia-Bellido and D.~G.~Figueroa,
  Phys.\ Rev.\ Lett.\  {\bf 98}, 061302 (2007);
J.~F.~Dufaux et al,
  Phys.\ Rev.\ D {\bf 76}, 123517 (2007).
\bibitem{GWdefects}
A.~Vilenkin,
  Phys.\ Lett.\ B {\bf 107}, 47 (1981); 
T.~Vachaspati and A.~Vilenkin,
  Phys.\ Rev.\ D {\bf 31}, 3052 (1985); 
S.~Olmez, V.~Mandic and X.~Siemens,
  Phys.\ Rev.\ D {\bf 81}, 104028 (2010)
  [arXiv:1004.0890 [astro-ph.CO]];  
D.~G.~Figueroa, M.~Hindmarsh and J.~Urrestilla,
  Phys.\  Rev.\  Lett.\  110, {\bf 101302} (2013)
  [arXiv:1212.5458 [astro-ph.CO]].
\bibitem{preheating}
  J.~H.~Traschen and R.~H.~Brandenberger,
  Phys.\ Rev.\ D {\bf 42}, 2491 (1990);
L.~Kofman, A.~D.~Linde and A.~A.~Starobinsky,
  Phys.\ Rev.\ Lett.\  {\bf 73}, 3195 (1994);
  Phys.\ Rev.\ D {\bf 56}, 3258 (1997).
\bibitem{MHz}
  A.~M.~Cruise, R.~M.~J.~Ingley and ,
  Class.\ Quant.\ Grav.\  {\bf 23}, 6185 (2006);
  T.~Akutsu, S.~Kawamura, A.~Nishizawa, K.~Arai, K.~Yamamoto, D.~Tatsumi, S.~Nagano and E.~Nishida {\it et al.},
  Phys.\ Rev.\ Lett.\  {\bf 101}, 101101 (2008)
  [arXiv:0803.4094 [gr-qc]].

\bibitem{LiddleLyth} 
  A.~R.~Liddle and D.~H.~Lyth,
  Cambridge, UK: Univ. Pr. (2000) 400 p

\bibitem{Suyama:2013dqa} 
  T.~Suyama and S.~Yokoyama,
  arXiv:1303.1254 [astro-ph.CO].

 \bibitem{Kofman2}   
P.~B.~Greene, L.~Kofman, A.~D.~Linde and A.~A.~Starobinsky,
  Phys.\ Rev.\ D {\bf 56}, 6175 (1997)
  [hep-ph/9705347].
\bibitem{prokopec}  
T.~Prokopec and T.~G.~Roos,
  Phys.\ Rev.\ D {\bf 55}, 3768 (1997)
  [hep-ph/9610400].

\bibitem{komatsu} E. Komatsu et al, arxiv 0804.1293, 2008.
\bibitem{CLUSTEREASY}
G.~N.~Felder,
  Comput.\ Phys.\ Commun.\  {\bf 179}, 604 (2008)
  [arXiv:0712.0813 [hep-ph]].

\bibitem{Figueroa1}
J.~Garc\'ia-Bellido, D.~G.~Figueroa, A. Sastre,
  Phys.\ Rev.\  D {\bf 77}, 043517 (2008);
\bibitem{Figueroa2}
 D.~G.~Figueroa, J.~Garcia-Bellido and A.~Rajantie,
  JCAP {\bf 1111}, 015 (2011).
\bibitem{rajantie} A.~Chambers and A.~Rajantie,
  Phys.\ Rev.\ Lett.\  {\bf 100}, 041302 (2008)
  [Erratum-ibid.\  {\bf 101}, 149903 (2008)]
  [arXiv:0710.4133 [astro-ph]];
  JCAP {\bf 0808}, 002 (2008)
  [arXiv:0805.4795 [astro-ph]].
\bibitem{Bond}   
J.~R.~Bond, A.~V.~Frolov, Z.~Huang and L.~Kofman,
  Phys.\ Rev.\ Lett.\  {\bf 103}, 071301 (2009)
  [arXiv:0903.3407 [astro-ph.CO]].
\bibitem{khlebnikovtkachev}
  S.~Y.~.Khlebnikov and I.~I.~Tkachev,
  Phys.\ Rev.\ Lett.\  {\bf 77}, 219 (1996)
  [hep-ph/9603378].
\bibitem{Chambers:2009ki} 
  A.~Chambers, S.~Nurmi and A.~Rajantie,
  JCAP {\bf 1001}, 012 (2010)
  [arXiv:0909.4535 [astro-ph.CO]].

\bibitem{Aad:2012tfa} 
  G.~Aad {\it et al.}  [ATLAS Collaboration],
  Phys.\ Lett.\ B {\bf 716}, 1 (2012)
  [arXiv:1207.7214 [hep-ex]];
  S.~Chatrchyan {\it et al.}  [CMS Collaboration],
  Phys.\ Lett.\ B {\bf 716}, 30 (2012)
  [arXiv:1207.7235 [hep-ex]].

\bibitem{Figueroa3}
J.~F.~Dufaux, D.~G.~Figueroa and J.~Garcia-Bellido,
  Phys.\ Rev.\ D {\bf 82}, 083518 (2010)
  [arXiv:1006.0217 [astro-ph.CO]].
\bibitem{Figueroa4}
K.~Enqvist, D.~G.~Figueroa and T.~Meriniemi,
  Phys.\ Rev.\ D {\bf 86}, 061301 (2012)
  [arXiv:1203.4943 [astro-ph.CO]].





\end{thebibliography}
\end{document}